\documentclass[twocolumn,prd,noshowpacs,nofootinbib,amsmath,amssymb,superscriptaddress]{revtex4}
\usepackage{graphicx,epsfig,psfrag,bm,amssymb}
\usepackage{dcolumn}
\usepackage{bm}
\usepackage{mciteplus}
\usepackage{color}
\usepackage{mathrsfs,amsmath, amsfonts,hepunits, color}
\usepackage{mciteplus}
\usepackage{tikz}
\usepackage{slashed}
\usetikzlibrary{arrows,shapes}
\usetikzlibrary{trees}
\usetikzlibrary{matrix,arrows} 				
\usepackage{slashed}

\newcommand{\be}{\begin{eqnarray}}
\newcommand{\ee}{\end{eqnarray}}
\def\lsim{\mathrel{\rlap{\lower4pt\hbox{\hskip 0.5 pt$\sim$}}
    \raise1pt\hbox{$<$}}}                
\def\gsim{\mathrel{\rlap{\lower4pt\hbox{\hskip1pt$\sim$}}
    \raise1pt\hbox{$>$}}}

\def\lsim{\mathrel{\rlap{\lower4pt\hbox{\hskip1pt$\sim$}}
    \raise1pt\hbox{$<$}}}
\def\gsim{\mathrel{\rlap{\lower4pt\hbox{\hskip1pt$\sim$}}
    \raise1pt\hbox{$>$}}}

\begin{document}

\title{ Probing New Physics with Underground Accelerators and Radioactive Sources}
\author{ Eder Izaguirre}
\author{ Gordan Krnjaic}
 \affiliation{Perimeter Institute for Theoretical Physics, Waterloo, Ontario, Canada    }
\author{ Maxim Pospelov}
 \affiliation{Perimeter Institute for Theoretical Physics, Waterloo, Ontario, Canada    }
 \affiliation{Department of Physics and Astronomy, University of Victoria,  
Victoria, British Columbia, Canada}

\begin{abstract}

New light, weakly coupled particles can be efficiently produced at existing and future high-intensity
 accelerators and radioactive sources in deep underground laboratories. Once produced, these particles 
can scatter or decay in large neutrino detectors (e.g Super-K and Borexino) housed in the same facilities.
 We discuss the production of weakly coupled scalars $\phi$ via nuclear de-excitation of an excited element into the
 ground state in two viable concrete reactions: the decay of the $0^+$ excited state of $^{16}$O populated via a $(p,\alpha)$ reaction on fluorine 
 and from radioactive $^{144}$Ce decay where the scalar is produced in the de-excitation of $^{144}$Nd$^*$, which occurs 
 along the decay chain. Subsequent scattering on electrons, $e(\phi,\gamma)e$, yields a mono-energetic signal that is observable 
 in neutrino detectors. We show that this proposed experimental set-up can cover new territory for masses $250\, {\rm keV}\leq m_\phi \leq 2 m_e$ 
 and couplings to protons and electrons, $10^{-11} \leq g_e g_p \leq 10^{-7}$. This parameter space 
 is motivated by explanations of the ``proton charge radius puzzle", thus this strategy adds a viable new physics component to the neutrino and nuclear astrophysics programs at underground facilities.

\end{abstract}

\maketitle

%
%

{\em Introduction.} In recent years, there has emerged a universal appreciation for new light, weakly-coupled 
degrees of freedom as generic possibilities for New Physics (NP) beyond Standard Model (SM). Considerable 
effort in ``intensity frontier" experiments is now devoted to NP searches  \cite{Essig:2013lka}.
In this {\it Letter} we argue that there is a powerful new possibility for probing these states by combining large underground
neutrino-detectors with either high luminosity underground accelerators or radioactive sources. 

Underground laboratories, typically located a few km underground, are shielded from most environmental backgrounds and 
 are ideal venues for studying rare processes such as low-rate nuclear reactions and solar neutrinos. 
 Thus far, these physics goals have been achieved with very different instruments: 
nuclear reactions relevant for astrophysics involve low-energy, high-intensity proton or ion beams colliding with 
fixed targets (such as the LUNA experiment at Gran Sasso), while solar neutrinos are detected with large 
volume ultra-clean liquid scintillator or water Cerenkov detectors (SNO, SNO+, Borexino, Super-K etc). 

\begin{figure}[t!]
 \vspace{0.cm}
 \hspace{-0.3cm}
\includegraphics[width=8.9cm]{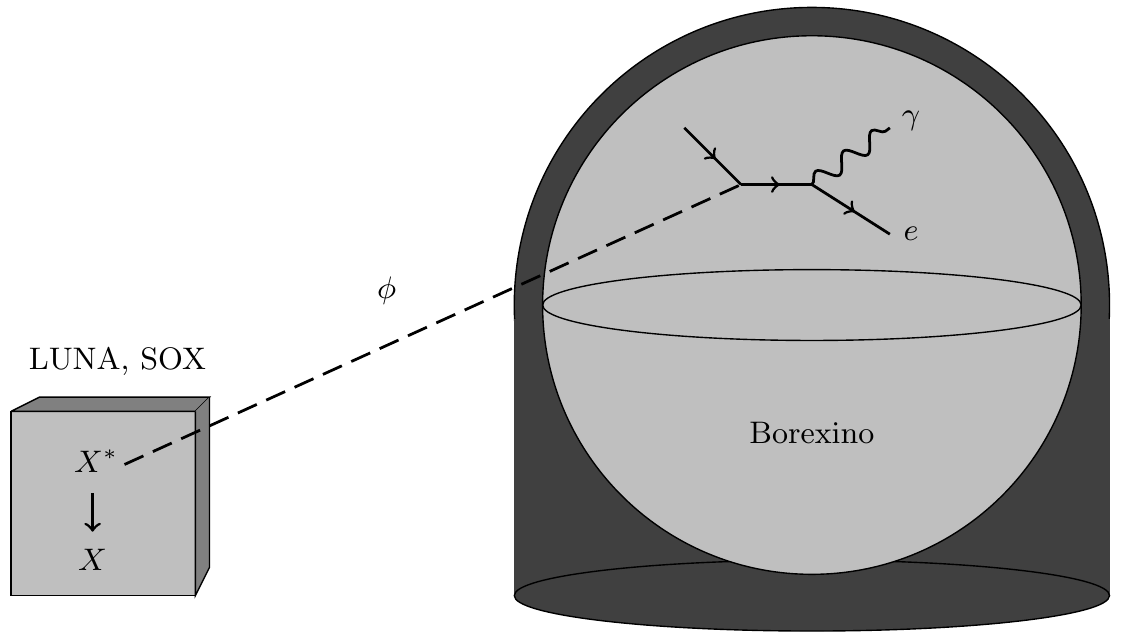}
  \caption{Schematic figure of $\phi$ production in
   a ``LUNA"-type underground accelerator via $p + $$^{19}{\rm F} \to  (^{16}{\rm O}^* \to$$~ ^{16}{\rm O} + \phi) + \alpha$ or
   a ``SOX''-type radioactive source via  $^{144}\text{Ce}-^{144}\text{Pr}(\bar\nu_e) \to \text{Nd}^* \to \text{Nd} + \phi$.
Subsequent detection at ``Borexino" proceeds via $\phi e \to e \gamma$ scalar conversion.}
\label{fig:schematic}
\end{figure}

In this {\it Letter} we outline a novel experimental strategy in which light, ``invisible" states $\phi$ are produced in 
underground accelerators or radioactive materials with $O({\rm MeV})$ energy release, 
and observed in nearby neutrino detectors in the same facilities as depicted in Fig. \ref{fig:schematic}:
\begin{eqnarray}
\label{eq:production}
X^* \to X + \phi,
~~&{\rm production~at~``LUNA" ~or~``SOX"}
\\
\label{eq:detection}
e+\phi\to e + \gamma, ~~
& {\rm detection~at~``Borexino"}.
\end{eqnarray}
Here $X^*$ is an excited state of element $X$, accessed via a nuclear reaction initiated by an underground accelerator (``LUNA'') or by a radioactive material (``SOX'')\footnote{Our idea is very generic, not specific to any single experiment or location, which is why quotation marks are used.}. In the ``LUNA"-type setup a proton beam collides against a fixed target, emitting a new light particle that travels unimpeded through the rock and scatters inside a ``Borexino"-type detector. Alternatively, in the ``SOX'' production scenario, 
designed
to study neutrino oscillations at short baselines, a radioactive material placed near a neutrino detector gives rise to the reaction in Eq.~\ref{eq:production} as an intermediate step of the radioactive material's decay chain.

We study one particularly well-motivated NP scenario with a $\lsim$ MeV scalar particle, very weakly $O(10^{-4})$ coupled 
to nucleons and electrons. This range of masses and couplings is not excluded by astrophysical or laboratory bounds, 
 and is motivated by the persistent proton charge-radius anomaly. 
Two concrete, viable possibilities for producing light scalars are considered:
\begin{itemize}
\item For the LUNA-type setup, we show that such light particles can be efficiently produced by populating the first excited 
6.05 MeV $0^+$ state of $^{16}$O in $(p,\alpha)$ reactions on fluorine.

\item For the SOX-type setup we find similarly powerful sensitivity from the $^{144}\text{Ce}-^{144}\text{Pr}(\bar\nu_e)$ radioactive source, which can produce a  
scalar with 2.19 or 1.49 MeV energies from the $^{144}\text{Nd}^*$ de-excitation that occurs along the decay chain.
\end{itemize}
The subsequent detection of a mono-energetic release in a Borexino-type detector with 6.05, 2.19, or 1.49 MeV will be free from substantial environmental backgrounds. The  strategy proposed in this {\it Letter} is capable of advancing the sensitivity to such states by many orders of magnitude, completely covering the parameter space relevant for the $r_p$ puzzle. 

{\em Scalar particles below {\rm 1 MeV}.}  New particles in the MeV and sub-MeV mass range are motivated 
by the recent $7\sigma$ discrepancy between the standard determinations of the proton charge radius, $r_p$, based on $e-p$ interactions
\cite{Mohr:2012tt}, and the recent, most precise determination of $r_p$ from the Lamb shift in muonic Hydrogen 
\cite{Pohl:2010zza,Antognini:1900ns}. One possible explanation for this anomaly is a new force between the electron(muon) and proton
\cite{TuckerSmith:2010ra,Barger:2010aj,Batell:2011qq} mediated by a $\sim$100 fm range force (scalar- or vector-mediated) that shifts the binding energies 
of Hydrogenic systems and  skews the determination of $r_p$.
Motivated by this anomaly, we consider a simple model with one light scalar $\phi$ that interacts with protons and
leptons,
\be
\label{L}
{\cal L}_\phi = \frac{1}{2} (\partial_\mu \phi)^2  -   \frac{1}{2} m_\phi^2 \phi^2 +  (g_p \bar p p + g_e \bar e e + g_\mu \bar \mu \mu ) \phi  ~,
\ee
and define $\epsilon^2 \equiv (g_e g_p)/e^2$. 
We  assume mass-weighted couplings to leptons, $g_e \propto (m_e/m_\mu) g_\mu$,
and no couplings to neutrons.  UV completing such a theory is challenging, so we regard this as a purely phenomenological 
model. The apparent corrections to the charge radius of the proton in regular and muonic hydrogen are \cite{TuckerSmith:2010ra,Barger:2010aj,Batell:2011qq}
\begin{eqnarray}
\label{eq:rp}
\left.\Delta r_p^2\right|_{e \rm H} = -\frac{6\epsilon^2}{m_\phi^2}~;~
\left.\Delta r_p^2\right|_{\mu \rm H} = -\frac{6\epsilon^2(g_\mu/g_e)}{m_\phi^2} f(a m_\phi)
\end{eqnarray} 
where $a \equiv (\alpha \,m_\mu m_p)^{-1}(m_\mu +m_p)$ is the $\mu$H Bohr radius and
$f(x) =  x^4(1+x)^{-4}$. Equating $\left.\Delta r_p^2\right|_{\mu \rm H}-\left.\Delta r_p^2\right|_{e \rm H}$
to the current discrepancy of $-0.063\pm0.009$~fm$^2$ \cite{Antognini:1900ns}, one obtains a relation between $m_\phi$ and $\epsilon$. Thus, for  
$m_\phi = 0.5$ MeV, the anomaly suggests  $\epsilon^2 \simeq 1.3\times 10^{-8}$.  For $m_\phi > 2m_e$, the $\phi\to e^+e^-$ 
process is highly constrained by searches for light Higgs bosons \cite{Essig:2013lka}, so
we consider the $m_\phi < 2 m_e$ region, which is relatively unconstrained. 
Since $g_e \ll g_p$, the $\phi-e$ coupling is suppressed relative to that of a massive photon-like particle, so
precision measurements of $\alpha$ and $(g-2)_e$ do not constrain this scenario. 


\begin{figure}[t!]
 \vspace{0.cm}
 \hspace{-0.65cm}
\includegraphics[width=9.cm]{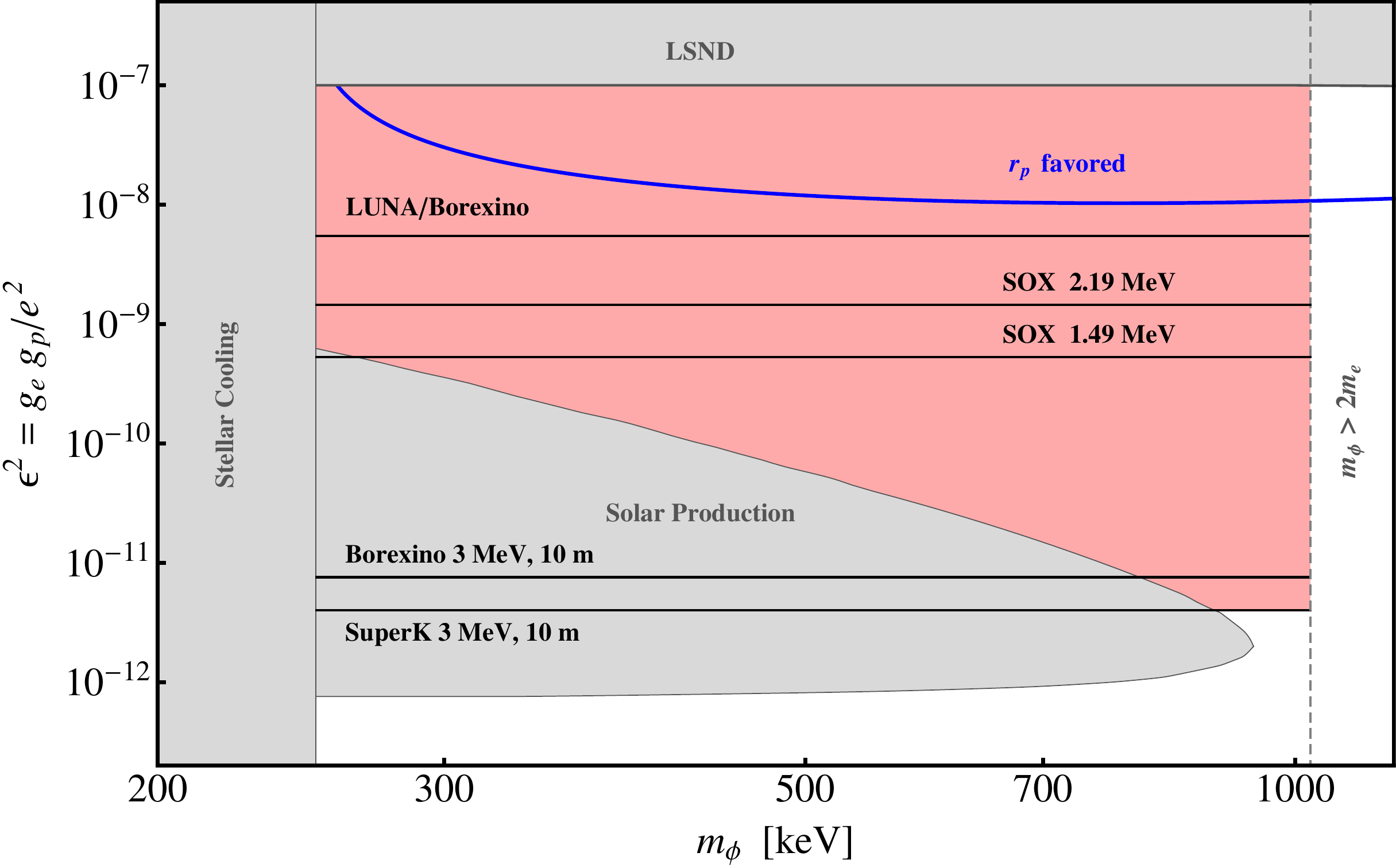}
\caption{Sensitivity projections for various experimental setups in terms
 of $\epsilon^2 = g_p g_e/e^2$ and $m_\phi$, which parametrize the NP 
explanation of the $r_p$ anomaly in Eq.~(\ref{eq:rp}); the blue band is the parameter space that resolves the puzzle.  
The  ``LUNA/Borexino" curve assumes a 400 keV proton beam with $10^{25}$ POT
 incident on a C$_3$F$_8$ target to induce $p + ^{19}$F $\to ({\rm ^{16}O}^*\to {\rm ^{16}O} +\phi) +  \alpha$ 
 reactions 100 m away from Borexino and yield 10 signal events  ($>3\sigma$) above backgrounds  \cite{Bellini:2012kz}. 
The  Borexino 3 MeV and SuperK 3 MeV lines assume the same setup with a 3 MeV $p$-accelerator 10 m away from 
each detector. The SuperK projection shows 100 signal events  ($>3\sigma$) above
 backgrounds at 6.05 MeV \cite{Mitsuda:2004gc}.  The SOX  lines assume a radioactive $^{144}\text{Ce}-^{144}\text{Pr}$ 
 source $7.15$ m away from Borexino with 50 and 165 events ($>3\sigma$) above backgrounds for 2.19 and 1.49 MeV lines respectively.
Shaded in gray are constraints from  solar production
  \cite{Bellini:2012kz}, LSND electron-neutrino scattering \cite{Auerbach:2001wg}, 
  and stellar cooling \cite{Raffelt:1994ry}, for which we assume $g_e = (m_e/m_p) g_p$.}
\label{fig:mainplot}
\end{figure}

The astrophysical and fixed-target constraints depend on the cross section for $e\phi\to e\gamma$ conversion, 
which for $m_\phi \ll m_e$ with a stationary electron target is
\be
\label{eq:dsigmaQ}
\frac{d\sigma}{dE} &=& 
\frac{\pi (g_e/e)^2 \alpha^2 (E- m_e)     }{ m_e Q^4 (Q - E + m_e)^2 } \biggl[
E (Q^2 - E Q - 2 m_e Q        \nonumber \\   && ~~~~   - 2m_e^2 ) + m_e (3 Q^2 + 3 Qm_e + 2m_e^2)  
   \biggr], 
\ee
where $E$ is the electron recoil energy and  $Q$ is the $\phi$ energy. 
At $Q\gg m_e$, this leads to a total cross section of 
\be
\label{sigma_num}
\sigma_{e\phi} \simeq \frac{\pi (g_e/e)^2 \alpha^2    }{ 2m_e Q} = 13~{\rm mbn}\times \frac{5 ~{\rm MeV}}{Q}\times \left(\frac{g_e}{e}\right)^2,
\ee
which determines the  in-medium $\phi$-absorption probability.
Absorption competes with the  $\phi \to \gamma \gamma$ decay, proceeding through loops of 
fermions $f$ with the width given by a standard formula, 
 \be
 \label{eq:width}
 \Gamma(\phi\to \gamma \gamma) =  \frac{  \,\alpha^2 \,m_\phi^3}{512\, \pi^3} \,\biggl|  \sum_f \frac{g_f }{m_f}N_c Q^2_f A_{1/2}(\tau_f)\biggl|^2~,
 \ee
  where $Q_f$ is the fermion charge, $\tau_f \equiv  m^2_\phi/4m_f^2$,  and
  \be
  A_{1/2}(\tau)= 2\tau^{-2} [\tau + (\tau - 1)\arcsin\sqrt{\tau} ].~
\ee 
An approximate proportionality to particle masses ensures that couplings to neutrinos are negligible. 

Processes (\ref{eq:dsigmaQ}), (\ref{eq:width}) define the gross features of $\phi$-phenomenology in cosmological and 
astrophysical settings. The ensuing constraints are summarized as follows: 
\begin{itemize}

\item Energy loss in stars via $e\gamma\to e\phi$  (red giants, white dwarfs etc) is exponentially suppressed for 
$m_\phi > T_{\rm star}$. This places a strong bound for $m_\phi \lsim 250$~keV, for the fiducial range of couplings.

\item The decay of $\phi$ in the early Universe at $T\sim m_\phi$ results in a negative shift of the ``effective number of neutrinos." For $m_\phi > 250$ keV 
the shift is moderate, $N_{\rm eff} \sim -0.5$ \cite{Nollett:2013pwa}, and 
can be easily compensated by the positive contributions from other light particles (e.g. sterile neutrinos).

\item SN physics: Low masses and sizable couplings, $g_{e,p} \sim 10^{-4}$, ensures the $\phi$ are trapped during the explosions, 
 and neither take energy from the explosive zones nor degrade the neutrino energies on account of $g_\nu=0$.

\item Emission of $\phi$ in solar nuclear reactions can be constrained using the Borexino search for solar axions \cite{Bellini:2012kz}, and 
 disfavors some fraction of the parameter space with $\epsilon^2$ in between $10^{-12}$ and $10^{-10}$, as shown in this work. 

\end{itemize}

In addition to astrophysical constraints, bounds on $\epsilon$ from direct searches of very light 
scalars typically probe $\epsilon^2 \gsim 10^{-7}$. When combined, existing constraints leave an unexplored part of the 
parameter space for the scalar model, $ 250~ {\rm keV} \lsim m_\phi < 2 m_e,~    10^{-10} \lsim  \epsilon^2 \lsim 10^{-7}$, 
and the $\Delta r_p$-motivated range falls in the middle of this allowed territory. The existing constraints are summarized in Fig. \ref{fig:mainplot}.




{\em Production of scalars in nuclear reactions.}
Searches of light scalar particles in nuclear reactions, such as $^3{\rm H} (p,\gamma) ^4{\rm He}$ and
$^{19}{\rm F}(p,\alpha)^{16}{\rm O}^*$
have been successfully implemented \cite{Freedman:1984sd,Kohler:1974zz} on the surface, where the main 
background comes from cosmic events.
For sub-MeV masses of $\phi$, the latter reaction is especially advantageous
as $\phi$ is produced in the  de-excitation  of the $0^+$ state:
\be
\label{eq:deexcitation-process}
 ^{16}{\rm O}^*(6.05)  \to ~ ^{16}{\rm O}  + \phi ~,
\ee
with energy release $Q = 6.05$ MeV. 
In the SM, the single-$\gamma$ decay of this state is not possible due to angular momentum conservation, and the main de-excitation 
process is $^{16}{\rm O}^*\to$  $^{16}{\rm O} +e^+ e^-$ with the long lifetime $ 96 \pm 7$ ps \cite{Tilley:1993zz}; thus, the relative branching 
 to new physics can be greatly enhanced. Following \cite{Snover:2003fg} for $m_\phi \ll Q$, the NP branching ratio 
 $\Gamma_{\phi}/\Gamma_{e^+e^-}$ is 
\be
{\cal B}r_\phi  = \frac{  8 \pi  (g_p/e)^2  Q^5}{   \alpha\, b(s) (Q - 2m_e)^3(Q + 2m_e)^2   }  \simeq 4 \times10^3  \left(\frac{g_p}{e}\right)^2  ,~~
\ee 
where $s = (Q-2m_e)/( Q+ 2m_e)$ and $b(s) \approx 0.92$ is defined in \cite{Snover:2003fg}.
The excited state $^{16}$O$^*$ can be efficiently produced in $\sim$ 100 keV--MeV $p$ accelerators.

To estimate the $\phi$ yield from $p + ^{19}\!{\rm F}~ \to ~^{16} \hspace{-0.1cm}~{\rm O}^*(6.05)+ \alpha ~,$
we model the cross section below 3 MeV
using \cite{Cuzzocrea1980,Angulo:1999zz}
and extrapolate to the Coulomb-suppressed region. Specifically, we take 
$\sigma(E) \simeq \sigma_0  f(E) $, 
with $\sigma_0 = 18$ mbn and model the Coulomb repulsion with
\be
 f(E < E_0) = \sqrt{\frac{E_0}{E}} \exp{\biggl( \sqrt{E_g/E_0} -\sqrt{E_g/E} \biggr)}  ~ ,
    \ee
in the $E < E_0 \equiv 1.5$ MeV range. 
Here $E_g  = 2 (\pi \alpha Z_F)^2 \mu  = 45.5$ MeV is the Gamow energy and $\mu$ is the proton-fluorine reduced 
mass, $E$ is the c.o.m. energy, and normalization ensures continuity at
 $f(E_0) = 1$, where repulsion can be neglected.

The signal yield for a proton beam of energy $E_p$ ({\em i.e.} the probability to produce a quantum of $\phi$ 
per each injected proton) and target material of Fluorine number-density $n_F$ is
\be
N_{\phi}(E_p) =  {\cal B}r_\phi \times n_F \int_{0}^{E_p} \!dE  \,\, \frac{\sigma_p(E)  }{|dE/dx|} ~~.
\ee
$|dE/dx|$ depends on the material that includes Fluorine, and is readily available in \cite{NIST}. 
For example, for the C$_3$F$_8$ material, the probability of producing one $\phi$ per injected proton is
$N_\phi(3~{\rm MeV}) \sim 3 \times 10^{-2}  (g_p/e)^2 $ .

The angular distribution of emerging $\phi$ is fully isotropic as nuclear recoil velocities are negligible, 
and the flux at the position of the detector is given by $\Phi_\phi = N_{\phi}(E_p) \times (dN_p/dt )/ 4\pi L^2$.
Inside the detector, the emitted $\phi$ scatter off electrons through $e\phi\to e\gamma$ with cross sections given by (\ref{eq:dsigmaQ}).
Thus, the only remaining free parameters (distance $L$, number of accelerated protons per second $dN_p/dt$, their energy $E_p$
as well as the number of electrons in the detector volume) are location, source, and detector-specific.

{\em Production of light states in radioactive decays.}  An alternative realistic mechanism for producing light weakly coupled 
particles is using the high-intensity radiative sources placed near a neutrino detector. In particular, we focus on the specific radioactive source $^{144}\text{Ce}-^{144}\text{Pr}(\bar\nu_e)$ motivated by 
the SOX proposal by the Borexino collaboration. The production of the scalar in this reaction proceeds via $^{144}\text{Ce} \to \beta \bar\nu+^{144}\text{Pr}$ followed by           
$^{144}\text{Pr}$$\to \beta \bar\nu + (^{144}\text{Nd}^* \to ^{144}\text{Nd}+ \phi)$. 
Once produced, the scalar can be detected at a neutrino detector.



{\em Possible accelerator realizations.}
All the ingredients for a successful realization of our idea currently exist at the underground Laboratori Nazionali del Gran Sasso (LNGS) in Italy, 
home of both the LUNA accelerator and Borexino detector.  In addition, there are several other facilities of interest
 including SNOLAB in Canada and the Kamioka Observatory in Japan. Both SNO+ and Super-K 
 detectors in these laboratories could be sensitive to new sub-MeV states if a proton accelerator were to be placed in their 
 vicinity. Furthermore, the Sanford Underground Research Facility (SURF) has current plans to host the Dual Ion Accelerators 
 for Nuclear Astrophysics (DIANA), which are expected to deliver 10-100 mA 3 MeV proton beams. SURF is 
 also home to the Large Underground Xenon (LUX) experiment, which despite its smaller volume compared to Borexino
  and Super-Kamiokande, could also be sensitive to new sub-MeV states.

The LUNA accelerator \cite{Broggini:2010mu} can deliver mA  currents of MeV scale proton energies  \cite{Montanari:2009zz}. Our main results and the plot with 
sensitivity projections assume a target which is not currently used by the LUNA experiment, (e.g. C$_3$F$_8$), but can easily be installed.
In Fig.~\ref{fig:mainplot} we show a realistic scenario assuming the existing 400 keV accelerator $L=100$ m away in the canonical
 LUNA scenario.  We also show projections for an upgraded 3 MeV beam \cite{Guglielmetti:2014rea}  10m away from the 
Borexino detector in the Gran Sasso service tunnel. For all our accelerator projections we optimistically assume $10^{25}$ 
protons-on-target (POT), achievable with a 50 mA beam 
running for one year.  Very importantly, at 6.05 MeV energy Borexino is almost background-free and has good energy resolution,
so that  even a handful of events $(\sim10)$  would show a significant excess in the corresponding energy bin, and constitute a discovery. 

One practical limitation of this proposal could be a requirement of not increasing the neutron background in LNGS.
In our example, the main source of neutrons is $\alpha$ nuclei produced in each reaction step, which yield neutrons 
in secondary collisions with target nuclei. 
Using \cite{Wrean:2000ni}, we estimate the neutron yield from $^{19}$F $(\alpha,n)$ $^{23}$Na in our setup
 to be $\sim O(\text{ few Hz})$.
Such low rates are irrelevant at LNGS, which can accommodate 10$^3$Hz, but might matter
 if alternate production methods are employed, thus requiring extra shielding.

The Super Kamiokande (SuperK) detector \cite{Fukuda:2002uc} in Kamioka, Japan 
contains a 50,000-ton water \v Cerenkov detector.
 In Fig.~\ref{fig:mainplot} we show
 the expected  $\epsilon$ sensitivity of a high-intensity  3 MeV proton source,
assuming a C$_3$F$_8 $ target 10 m away from the detector.
Despite a penalty due to a relatively high threshold for the electron energy in SuperK, one can see an incredibly strong potential 
for the reach to new physics. 

{\em Possible radioactive source realizations.}
For scalar production via radioactive decays, one possibility is phase B of the SOX proposal by the Borexino collaboration 
\cite{Borexino:2013xxa}, which intends to deploy a $\sim 2$ PBq 
source of $^{144}$Ce-$^{144}$Pr  7.15 m from the Borexino center.
Roughly  2\% of $^{144}$Ce decays are accompanied by the $\gamma$-radiation
from the decay of the metastable Nd$^*$ daughter nuclei described above. 
The $1.49$ and $2.19$ MeV transition energies are well above the Borexino threshold, so this
method covers the full mass range of interest, generating 
 $\sim 10^{13}  (g_p/e)^2 $ $\phi$-particles per second.
 Given the planned exposures \cite{Borexino:2013xxa}, we estimate the Borexino reach in this case, and add corresponding 
sensitivity lines on Fig. 2. 




{\em Existing constraints.}  
While many of the past beam-dump experiments can be sensitive to sub-MeV particles, we concentrate on the one that is able to constrain
the product of $g_pg_e$, namely the LSND experiment at Los Alamos. 
Its measurement of the elastic electron-neutrino  cross section \cite{Auerbach:2001wg} is also
 sensitive to light scalars that induce $e\gamma$ events due to scattering on electrons.
 This analysis has previously been used to constrain new vector particles produced in $\pi^0$ decays to dark sector states \cite{Batell:2009di,deNiverville:2011it}. 
 In our scenario, a scalar $\phi$ cannot be produced from pseudoscalar $\pi^0$ decays. Instead, the 
dominant process is  $\pi^-$ absorption via $\pi^- p \to n \phi$. The 
  analogous SM process $\pi^-  p \to n \gamma$ has branching ratio $\sim 35 \%$ \cite{Samios1960}, so we approximate the $\phi$ 
  branching as $\sim \epsilon^2 \times 35 \%$. Taking the $\pi^-$ production rate at LSND to be 
 roughly 10\% of the $\pi^+$ production  implies $\sim 10^{22}\, \pi^-$ for  the exposure in
   \cite{Auerbach:2001wg}. Assuming isotropic $\phi$ emission and  
  the scattering cross section  in Eq.~(\ref{eq:dsigmaQ}) with $Q\to m_p + m_{\pi^-} - m_n \simeq m_\pi$, and implementing 
the cuts from this analysis,  we obtain a roughly flat bound $\epsilon^2 \lsim 10^{-8}$ for $m_\phi < $ MeV as shown in Fig.~\ref{fig:mainplot}.
     This sensitivity exceeds even the bounds from $(g-2)_e$ from \cite{Giudice:2012ms}, which only imply $\epsilon^2 \lsim 10^{-7}$ over
     this mass range, assuming mass weighted couplings $g_p =(m_p/m_e) g_e$; for $g_e = g_p$, the bounds from $(g-2)_e$ are comparable to those set 
     by LSND. 

In the 100 keV -- MeV mass window $\phi$'s cannot be produced thermally in the solar interior, but can be 
produced in nuclear reactions. A particularly relevant process is $p+d \to ^3\!\!{\rm He} + \phi$ (that accompanies 
the $d(p,\gamma)^3{\rm He}$ reaction occurring for every 
individual $pp$ event of energy generation). If $\phi$ is sufficiently long lived, and not absorbed in the 
solar interior, it will reach the Earth and deposit 5.5 MeV of energy in Borexino. 
The absence of such events  \cite{Bellini:2012kz} sets an important constraint on our model. 
 
 The solar flux of 5.5 MeV $\phi$ particles at Borexino is approximated using the $pp$-neutrino flux  via 
 \be
\Phi_{\phi,{\rm solar}}  \simeq \epsilon^2 P_{\rm esc} P_{\rm surv} \Phi_{pp\nu}~, 
 \ee
 where $\Phi_{pp\nu} = 6.0 \times 10^{10} \cm^{-2}$ s$^{-1}$ \cite{Bellini:2012kz}.
 The probability of escaping the sun is $P_{\rm esc} = \exp(- \int^{R_\odot} dr n_\odot \sigma_{e\phi})$, 
 the probability that the scalar does not decay between the Sun and the Earth is $P_{\rm surv} = \exp(-\ell_\odot/\ell_{\phi})$,
where  $\ell_\phi =  Q c/ m_\phi \Gamma(\phi\to \gamma \gamma)$ is 
 the boosted decay length, and $\ell_\odot$  is the Earth-Sun distance. 
 The Borexino rate is 
 \be
 \dot N_{\phi e} &=& \Phi_{\phi,{\rm solar}}  \, n_B \sigma_{e\phi} \, V_B 
 \ee
 where  $n_{\odot,B}$ are mean-solar and Borexino $e^-$ densities, $V_B$ is the Borexino volume, and the 
 cross section off electrons is given in (\ref{sigma_num}). 
The current limits on this  process are $O(5)$ events \cite{Bellini:2012kz} and the constraint  
is depicted by the oval region in Fig.~\ref{fig:mainplot}. For $\epsilon^2 \gsim 10^{-10}$,  scattering
off electrons prevents $\phi$ from leaving the Sun and for $\epsilon^2 \lsim  \times 10^{-12}$ the production 
and scattering are insufficient to yield an appreciable signal at Borexino.

The constraints from thermal energy loss in red giants and white dwarfs 
follow the standard considerations. Calculating the thermal energy loss $\propto g_e^2 \exp(-m_\phi/T_{\rm star})$ and reinterpreting the axion
constraints from \cite{Raffelt:1994ry}, we exclude the $m_\phi \lsim 250$ keV parameter space for all $\epsilon$ of interest. 




To conclude, in this {\it Letter} we have proposed a novel strategy to hunt for sub-MeV particles produced in underground accelerators and radioactive
sources located 10 - 100 m away from large underground neutrino detectors. 
This experimental program offers unprecedented sensitivity to a variety of NP scenarios including those that resolve the $r_p$ puzzle.

{\em Acknowledgments.}
We thank Drs. A. Arvanitaki, J. Beacom, and I. Yavin for helpful conversations.
The Perimeter Institute for Theoretical Physics is supported by the Government of Canada through Industry Canada
and by the Province of Ontario.

\bibliographystyle{apsrevM}
\bibliography{UndergroundAccel}

\end{document}